\documentclass[twocolumn,pre,showpacs]{revtex4}

\usepackage{graphicx,amsmath,amssymb}
\usepackage{color}

\newcommand\defn{\textit}
\newcommand\mat{\mathbf}
\renewcommand\vec{\mathbf}
\newcommand\eref[1]{(\ref{#1})}
\newcommand\av[1]{\langle#1\rangle}
\newcommand\etal{\textit{et~al.}}

\begin{document}
\title{Vertex similarity in networks}

\author{E. A. Leicht, Petter Holme, and M. E. J. Newman}
\affiliation{Department of Physics, University of Michigan, Ann Arbor,
MI 48109, U.S.A.}

\begin{abstract}
We consider methods for quantifying the similarity of vertices in networks.
We propose a measure of similarity based on the concept that two vertices
are similar if their immediate neighbors in the network are themselves
similar.  This leads to a self-consistent matrix formulation of similarity
that can be evaluated iteratively using only a knowledge of the adjacency
matrix of the network.  We test our similarity measure on
computer-generated networks for which the expected results are known, and
on a number of real-world networks.
\end{abstract}


\maketitle

\section{Introduction and background}
The study of networked systems, including computer networks, social
networks, biological networks, and others, has attracted considerable
attention in the recent physics literature~\cite{AB02,DM02,Newman03d}.  A
number of structural properties of networks have been the subject of
particularly intense scrutiny, including the lengths of paths between
vertices~\cite{WS98,CL02b,CH03,FFH04}, degree
distributions~\cite{BA99b,DMS00,KRL00}, community
structure~\cite{GN02,Radicchi04,GSA04,DM04}, and various measures of vertex
centrality~\cite{WS98,Newman01c,GKK01b,Holme02a}.

Another important network concept that has received comparatively little
attention is vertex similarity.  There are many situations in which it
would be useful to be able to answer questions such as ``How similar are
these two vertices?'' or ``Which other vertices are most similar to this
vertex?''  Of course, there are many senses in which two vertices can be
similar.  In the network of the World Wide Web, for instance, in which
vertices represent Web pages, two pages might be considered similar if the
text appearing on them contains many of the same words.  In a social
network representing friendship between individuals, two people might be
considered similar if they have similar professions, interests, or
backgrounds.  In this paper we consider ways of determining vertex
similarity based solely on the structure of a network.  Given only the
pattern of edges between vertices in a network, we ask, can we define
useful measures that tell us when two vertices are similar?  Similarity of
this type is sometimes called \defn{structural similarity}, to distinguish
it from social similar, textual similarity, or other similarity types.  It
is a basic premise of research on networks that the structure of a network
reflects real information about the vertices the network connects, so it
appears reasonable that meaningful structural similarity measures might
exist.  Here we show that indeed they do and that they can return useful
information about networks.

The problem of quantifying similarity between vertices in a network is not
a new one.  The most common approach in previous work has been to focus on
so-called \defn{structural equivalence}.  Two vertices are considered
structurally equivalent if they share many of the same network neighbors.
For instance, it seems reasonable to conclude that two individuals in a
social network have something in common if they share many of the same
friends.  Let $\Gamma_i$ be the neighborhood of vertex $i$ in a network,
i.e., the set of vertices that are directly connected to $i$ via an edge.
Then the number of common friends of $i$ and $j$ is
\begin{equation}
\sigma_\mathrm{unnorm} = |\Gamma_i\cap\Gamma_j|,
\end{equation}
where $|x|$ indicates the cardinality (i.e., number of elements in) the
set $x$, so that $|\Gamma_i|$, for instance, is simply equal to the degree
of vertex $i$.

The quantity $\sigma_\mathrm{unnorm}$ can be regarded as a rudimentary
measure of similarity between $i$ and $j$.  It is, however, not entirely
satisfactory.  It can take large values for vertices with high degree even
if only a small fraction of their neighbors are the same, and in many cases
this runs contrary to our intuition about what constitutes similarity.
Commonly therefore one normalizes in some way---for instance so that the
similarity is one when~$\Gamma_i=\Gamma_j$.  We are aware of at least three
previously-proposed ways of doing this~\cite{Jaccard01,Salton89,rav:meta}:
\begin{subequations}
\begin{eqnarray}
\sigma_\mathrm{Jaccard} &=& \frac{|\Gamma_i\cap\Gamma_j|}
                            {|\Gamma_i\cup\Gamma_j|},\\
\label{eq:structequivb}
\sigma_\mathrm{cosine}  &=& \frac{|\Gamma_i\cap\Gamma_j|}
                            {\sqrt{|\Gamma_i|\,|\Gamma_j|}},\\
\sigma_\mathrm{min}     &=& \frac{|\Gamma_i\cap\Gamma_j|}
                            {\min(|\Gamma_i|\,|\Gamma_j|)}.
\end{eqnarray}
\label{eq:structequiv}
\end{subequations}
The first of these, commonly called the \defn{Jaccard index}, was proposed
by Jaccard over a hundred years ago~\cite{Jaccard01}; the second, called
the \defn{cosine similarity}, was proposed by Salton in 1983 and has a long
history of study in the literature on citation
networks~\cite{SM83,Salton89,Hamers89}.  (Measures nonlinear in
$\sigma_\mathrm{unnorm}$ are also possible.  For example,
Refs.~\cite{burt:se} and~\cite{gold:asse} propose measures involving
$\sqrt{\sigma_\mathrm{unnorm}}$ and~$\sigma_\mathrm{unnorm}^2$,
respectively.)

There are, however, many cases in which vertices occupy similar structural
positions in networks without having common neighbors.  For instance, two
store clerks in different towns occupy similar social positions by virtue
of their numerous professional interactions with customers, although it is
quite likely that they have none of those customers in common.  Two CEOs of
companies occupy similar positions by virtue of their contacts with other
high-ranking officers of the companies for which they work, although again
none of the individual officers need be common to both companies.
Considerations of this kind lead us to an extended definition of network
similarity known as \defn{regular equivalence}.  In this case vertices are
said to be similar if they are connected to other vertices \emph{that are
themselves similar}.  It is upon this idea that the measures developed in
this paper are based.

Regular equivalence is clearly a self-referential concept: one needs to
know the similarity of the neighbors of two vertices before one can compute
the similarity of the two vertices themselves.  It comes as no surprise to
learn, therefore, that traditional algorithms for computing regular
equivalence have an iterative or recursive nature.  Two of the best known
such algorithms, REGE and CATREGE~\cite{boev:rege}, proceed by searching
for optimal matching between the neighbors of the two vertices, while other
authors have formulated the calculation as a optimization
problem~\cite{bata:rege}.

\begin{figure}
\includegraphics[width=3.5cm]{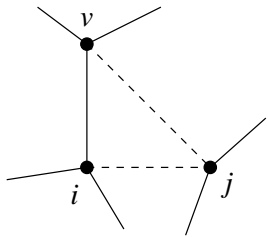}
\caption{A vertex $j$ is similar to vertex $i$ (dashed line) if $i$ has a
network neighbor $v$ (solid line) that is itself similar to $j$.}
\label{fig:simdef}
\end{figure}

In this paper, we take a different approach, constructing measures of
similarity using the methods of linear algebra.  The fundamental statement
of our approach is that vertices $i$ and $j$ are similar if either of them
has a neighbor $v$ that is similar to the other---see
Fig.~\ref{fig:simdef}.  Coupled with the additional assumption that
vertices are trivially similar to themselves, this gives, as we will see, a
sensible and straightforward formulation of the concept of regular
equivalence for undirected networks.  The method has substantial advantages
over other similarity measures: it is global---unlike the Jaccard index and
related measures, it depends on the whole graph and allows vertices to be
similar without sharing neighbors; it has a transparent theoretical
rationale, which more complex methods like REGE and CATREGE
lack~\cite{boev:rege}; it avoids the convergence problems that have plagued
optimization methods; and it is comparatively fast, since its
implementation can take advantage of standard, hardware optimized, linear
algebra software.

Some previous authors have also considered similarity measures based on
matrix methods~\cite{JW02,blondel:sim}.  We discuss the differences
between our measure and these previous ones in Section~\ref{sect:previous}.

This paper is organized as follows: In Section~\ref{sect:definition} we
present the derivation of our structural similarity measure.
Section~\ref{sect:tests} we test the measure on a number of networks,
including computer-generated graphs (Sections~\ref{sect:stratified}
and~\ref{sect:alpha}) and real-world examples
(Sections~\ref{sect:thesaurus} and~\ref{sect:friendship}).  In
Section~\ref{sect:concs} we give our conclusions.

\section{A measure of similarity}
\label{sect:definition}
The fundamental principle behind our measure of structural similarity in
networks is that $i$ is similar to $j$ if $i$ has a network neighbor $v$
that is itself similar to $j$ (Fig.~\ref{fig:simdef}).  Alternatively,
swapping $i$ and $j$, we could say that $j$ is similar to $i$ if it has a
neighbor $v$ that is similar to $i$.  Despite the apparent asymmetry
between $i$ and $j$ in these statements we will see that they both lead to
the same similarity measure, which is perfectly symmetric.

Our definition of similarity is clearly recursive and hence we need to
provide some starting point for the recursion in order to make the results
converge to a useful limit.  The starting point we choose is to make each
vertex similar to itself, which is natural in most situations.  Our
definition of similarity will thus have two components: the neighbor term
of the previous paragraph and the self-similarity.

Thus our first guess at the form of the similarity (we will improve it
later) is to write the similarity $S_{ij}$ of vertex $i$ to vertex $j$ as
\begin{equation}
S_{ij} = \phi \sum_v A_{iv}S_{vj} + \psi\delta_{ij},
\label{eq:initialSimExp}
\end{equation}
where $A_{ij}$ is an element of the adjacency matrix of the network taking
the value
\begin{equation}
A_{ij} = \left\lbrace\begin{array}{ll}
           1 & \qquad\mbox{if there is an edge between $i$ and $j$,} \\
           0 & \qquad\mbox{otherwise,}
         \end{array}\right.
\end{equation}
and $\phi$ and $\psi$ are free parameters whose values control the balance
between the two components of the similarity.

Considering $S_{ij}$ to be the $ij$ element of a similarity
matrix $\mat{S}$, we can write Eq.~\eref{eq:initialSimExp} in matrix form
as
\begin{equation}
\mat{S} = \phi\mat{A}\mat{S} + \psi\mat{I},
\label{eq:matform}
\end{equation}
where $\mat{I}$ is the identity matrix.  Rearranging, this can also be
written $\mat{S} = \psi [ \mat{I} - \phi\mat{A} ]^{-1}$.  As we see, the
parameter $\psi$ merely contributes an overall multiplicative factor to our
similarity.  Since in essentially all cases we will be concerned not with
the absolute magnitude of the similarity but only with the relative
similarity of different pairs of vertices, we can safely set $\psi=1$,
eliminating one of our free parameters, and giving
\begin{equation}
\mat{S} = [ \mat{I} - \phi\mat{A} ]^{-1}.
\label{eq:matsim}
\end{equation}
This expression for similarity bears a close relation to the matrix-based
centrality measure of Katz~\cite{katz:cent}.  In fact, the Katz centrality
of a vertex is equal simply to the sum of that vertex's similarities to
every other vertex.  This is a natural concept: a vertex is prominent in a
network if it is closely allied with many other vertices.

We can also consider the similarity of $i$ and $j$ when $j$ has a
neighbor $v$ that is similar to $i$.  In that case,
\begin{equation}
S_{ij} = \phi \sum_v S_{iv}A_{vj} + \psi\delta_{ij}.
\end{equation}
It is trivial to show however that this leads to precisely the same
expression, Eq.~\eref{eq:matsim}, for the similarity in the end.  Thus our
definition provides only one similarity value for any pair of vertices,
given by the symmetric matrix $\mat{S}$ of Eq.~\eref{eq:matsim}.

The remaining parameter $\phi$ in Eq.~\eref{eq:matsim} is still free.  To
shed light on the appropriate value for this parameter, let us expand the
similarity as a power series thus:
\begin{equation}
\mat{S} = \mat{I} + \phi\mat{A} + \phi^2\mat{A^2} + \ldots
\label{eq:series}
\end{equation}
Noting that the element $\bigl[\mat{A}^l\bigr]_{ij}$ is equal to the number
of (possibly self-intersecting) network paths of length $l$ from $i$
to $j$, this equation gives us an alternative, term-by-term interpretation
of our similarity measure.  The first term says that a vertex is
identically similar to itself.  The second term says that vertices that are
immediate neighbors of one another have similarity $\phi$.  The third term
says that vertices that are distance two apart on the network have
similarity $\phi^2$.  And so forth.

But notice also that vertices that have many paths of a given length are
considered more similar than those that have few.  The similarity of
vertices $i$ and $j$ acquires a contribution $\phi^2$ for \emph{every} path
of length 2 from $i$ to~$j$.  We note however that some pairs of vertices
are \emph{expected} to have one or even many such paths between them:
vertices with very high degree, for instance, will almost certainly have
one or several paths of length two connecting them, even if connections
between vertices are just made at random.  So simple counts of number of
paths are not enough to establish similarity.  We need to know when a pair
of vertices has more paths of a given length between than we would expect
by chance.

This suggests a strategy for choosing~$\phi$.  We will normalize each term
in our series by dividing the number of paths of length~$l$ (given by the
power of the adjacency matrix) by the \emph{expected} number of such paths,
were vertices in the network connected at random.  Then each term will be
greater or less than unity by a factor representing the extent to which the
corresponding vertices have more or fewer paths of the appropriate length
than would be expected by chance.  In fact, there is no single choice of
the parameter $\phi$ that will simultaneously achieve this normalization
for every term in the series but, as we will show, there is a choice that
achieves it approximately for every term, and exactly in the asymptotic
limit of high terms in the series, if we allow a slight (and with hindsight
sensible) modification of Eq.~\eref{eq:matsim}.

\subsection{Expected number of paths}
Let us generalize the series, Eq.~\eref{eq:series}, to allow an
independent coefficient for each term and for each vertex pair~$i,j$:
\begin{equation}
S_{ij} = \sum_{l=0}^\infty C_l^{ij} \bigl[ \mat{A}^l \bigr]_{ij}.
\label{eq:simsum}
\end{equation}
And let us choose (for the moment) each coefficient to be equal to 1 over
the expected number of paths of the corresponding length between the same
pair of vertices on a network with the same degree sequence as the network
under consideration, but in which the vertices are otherwise randomly
connected.  Such a network is called a \defn{configuration model}, and the
configuration model has been widely studied in the networks
literature~\cite{Luczak92,MR95,NSW01}.

The zeroth-order coefficient $C_0^{ij}$ is trivial: there are no paths of
length zero between vertices $i$ and $j$ unless $i=j$, in which case there
is exactly one such path.  So $C_0^{ij}=\delta_{ij}$.  The first-order term
is more interesting.  If vertices $i$ and $j$ have degrees $k_i$ and $k_j$
respectively, then we can calculate the expected number of paths of length
one between them as follows.  For any of the $k_i$ edges emerging from
vertex~$i$, there are $2m$ places where it could terminate, where $m$ is
the total number of edges in the network.  Of these, $k_j$~end at vertex
$j$ and hence result in a direct path of length one from $i$ to~$j$.  Thus
for each edge emerging from $i$ there is a probability $k_j/2m$ of a
length-one path to $j$, and overall the expected number of such paths
is~$k_ik_j/2m$.  Thus
\begin{equation}
C_1^{ij} = {2m\over k_ik_j}.
\label{eq:valc1}
\end{equation}

Now consider the second-order term in the series.  A path of length two
between $i$ and $j$ must go through a single intermediate vertex $v$, whose
degree we denote $k_v$.  Using the argument of the preceding paragraph,
the expected number of paths of length one from $i$ to $v$ is $k_ik_v/2m$.
This uses up one of the edges emerging from $v$, leaving $k_v-1$ remaining
edges and thus the expected number of paths of length one from $v$ to $j$,
given that there is already a path from $i$ to $v$, is $(k_v-1)k_j/2m$.
The expected number of paths of length two from $i$ to $j$ via $v$ is then
the product $k_ik_v(k_v-1)k_j/(2m)^2$.  Summing over all $v$, the total
expected number of paths of length two is
\begin{equation}
{k_ik_j\over(2m)^2} \sum_v k_v(k_v-1)
   = {k_ik_j\over2m} \left( {\av{k^2}-\av{k}\over\av{k}} \right),
\label{eq:pathsl2}
\end{equation}
where $\av{k}$ and $\av{k^2}$ are the mean degree and mean-square degree of
the network, and we have made use of the result $2m=n\av{k}$, where $n$ is
the total number of vertices in the network.

\begin{figure}
\includegraphics[width=\columnwidth]{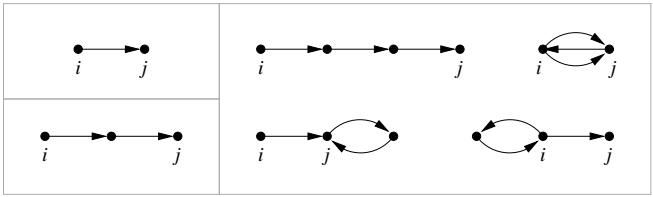}
\caption{There is only one possibility topology for paths of length one
between distinct vertices, and only one for paths of length two, but there
are four possible topologies for paths of length three.}
\label{fig:topo}
\end{figure}

For paths of length three and greater, the calculations become more
complicated.  Since paths can be self-intersecting, we have to consider
topologies for those paths that include loops or that traverse the same
edge more than once.  While there is only one topology for paths of length
one or two between a specified pair of vertices, there are four distinct
topologies for paths of length three (Fig.~\ref{fig:topo}), and we must
calculate and sum the expected numbers of each of them to get the total
expected number of paths.  The end result for paths of length three is
\begin{equation}
\frac{k_ik_j}{2m}\left[\left(\frac{\langle
      k^2\rangle-\langle k\rangle}{\langle k\rangle} \right)^{2} + k_i + k_j - 1 \right].
\label{eq:pathsl3}
\end{equation}

\begin{figure}
\resizebox{\linewidth}{!}{\includegraphics{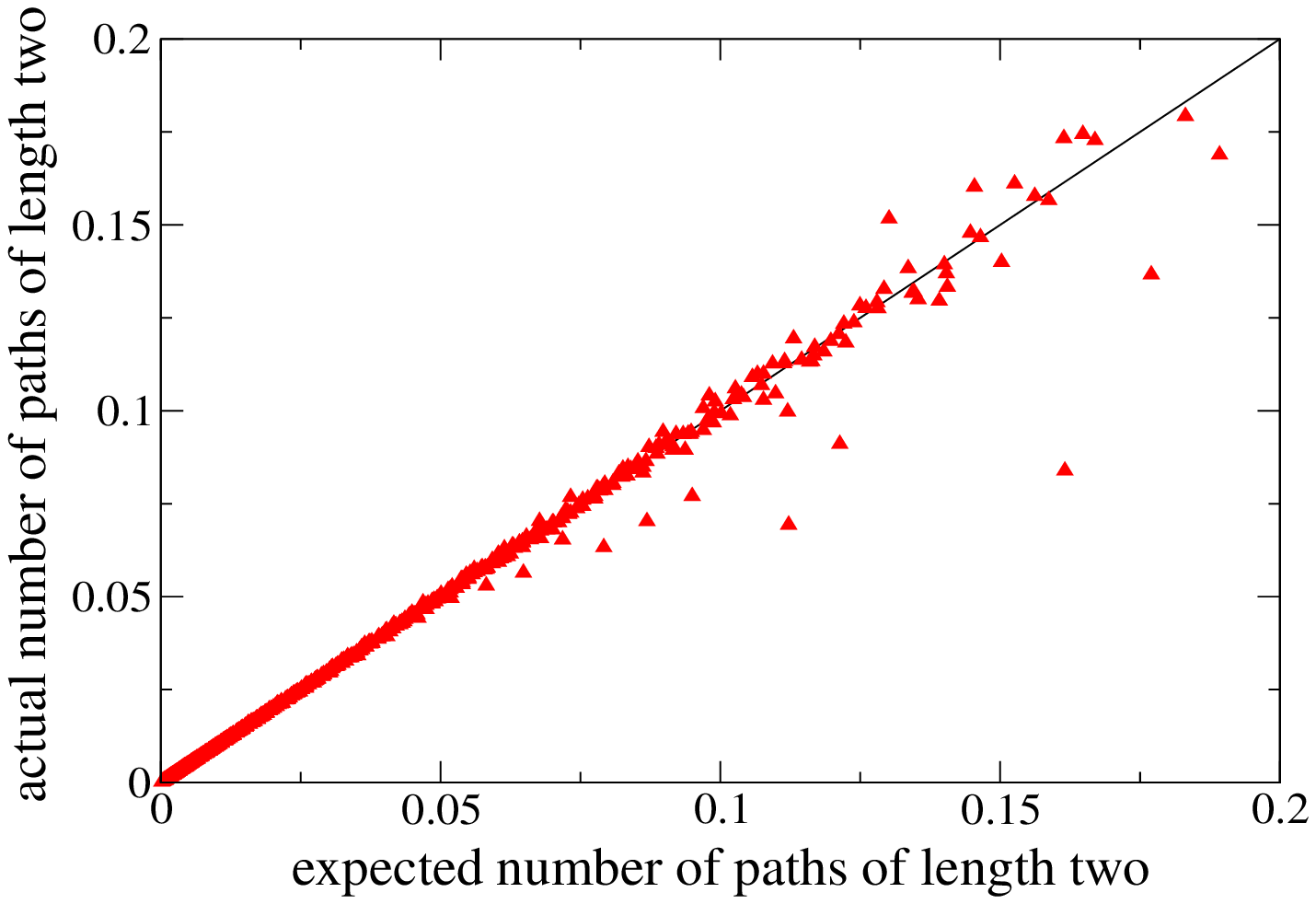}}\\
\resizebox{\linewidth}{!}{\includegraphics{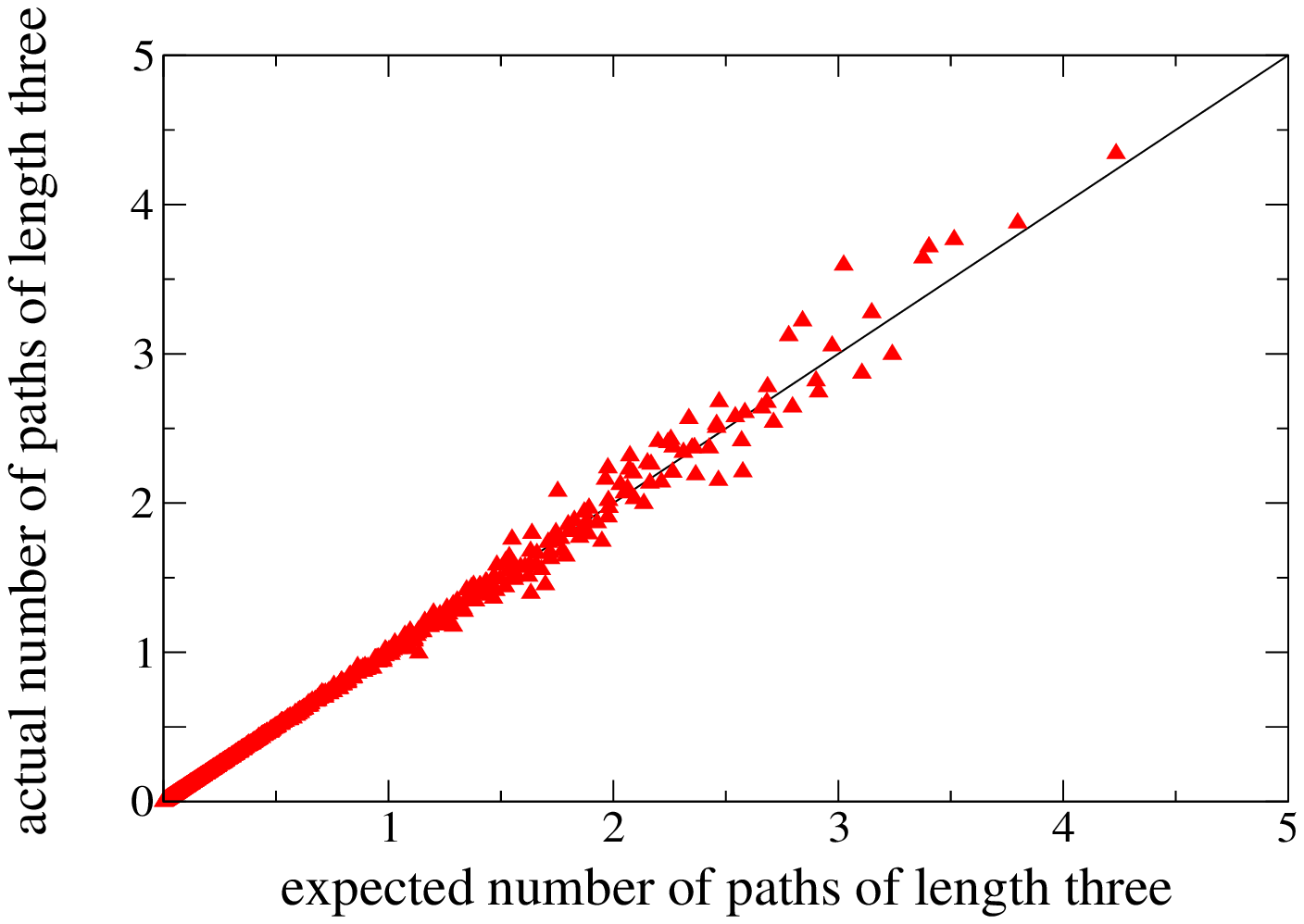}}\\
\caption{(a)~The actual number of paths of length two between vertex
  pairs in a configuration model versus the expected number of paths, as
  determined from Eq.~\eref{eq:pathsl2}. (b)~Same as in plot~(a), but for
  paths of length three.}
\label{fig:shortpaths}
\end{figure}

As a check on our calculations, we compare our analytic expressions for the
numbers of paths of length 2 and 3 to actual path counts for randomly
generated networks in Fig.~\ref{fig:shortpaths}.  There is increased
scatter in the numerical data at longer path lengths because the graphs
studied are finite in size, but overall the agreement between analytic and
numerical calculations is good.

While this is rewarding, it is not possible to extend this line of
investigation much further.  The expressions for expected numbers of paths
become more complicated as path length increases and the numbers of
distinct topologies multiply.  So instead, we take a slightly different
approach.

The expected number of paths of length $l$ from $i$ to $j$ can be written
as the $j$th element of the vector $\vec{p}_l$ given by
\begin{equation}
\vec{p}_l = \mat{A}^l \vec{v},
\label{eq:eigenvector}
\end{equation}
where the vector $\vec{v}$ has all elements zero except for $v_i=1$.  In
the limit of large $l$, the vector $\vec{p}_l$ tends toward (a multiple of)
the leading eigenvector of the adjacency matrix, and hence in this limit we
have $\vec{p}_{l+1}=\lambda_1\vec{p}_l$, where $\lambda_1$ is the largest
eigenvalue of $\mat{A}$.  Thus the number of paths from $i$ to $j$
increases by a factor of $\lambda_1$ each time we add one extra step to the
path length.  The first step of the path violates this rule: we know the
number of paths increases by exactly a factor of $k_i$ on the first step.
Furthermore, since our paths are constrained to end at vertex $j$, the last
step must end at one of the $k_j$ edges emanating from $j$, out of a total
of $2m$ possible places that it could end.  This introduces a factor of
$k_j/2m$ into the expected number of paths.  Thus, to within a
multiplicative constant, the number of paths of length $l$ from $i$ to $j$,
for large $l$, should be $(k_ik_j/2m)
\lambda_1^{l-1}$.

This expression is not in general correct for small $l$.  It is however
correct for the particular case $l=1$ of paths of length one (see
Eq.~\eref{eq:valc1}) and we expect it to be approximately correct for other
intermediate values of $l>1$.  Guided by these results, we therefore choose
the constants $C_l^{ij}$ appearing in Eq.~\eref{eq:simsum} to take the
values:
\begin{equation}\label{eq:cldef}
C_l^{ij} = {2m\over k_ik_j}\,\lambda_1^{-l+1},
\end{equation}
for $l\ge1$, with $C_0^{ij}=\delta_{ij}$.  These values approximate the
desired values based on expected numbers of paths and are asymptotically
correct in the limit of large~$l$.

\subsection{Derivation of the similarity}
There is one more issue we need to deal with with before we arrive at a
final expression for our similarity.  If we simply substitute $C_l^{ij}$
from Eq.~\eref{eq:cldef} into Eq.~\eref{eq:simsum} we produce a series that
unfortunately does not converge.  Thus, to ensure convergence, we introduce
an extra numerical factor~$\alpha$, giving a series thus:
\begin{eqnarray}
S_{ij} &=& \delta_{ij} + {2m\over k_ik_j}
  \sum_{l=1}^\infty \alpha^l \lambda_1^{-l+1} \bigl[ \mat{A}^l \bigr]_{ij}
  \nonumber\\
  &=& \biggl[ 1 - {2m\lambda_1\over k_ik_j} \biggr] \delta_{ij} +
    {2m\lambda_1\over k_ik_j} \biggl[ \biggl(
    \mat{I} - {\alpha\over\lambda_1} \mat{A} \biggr)^{-1} \biggl]_{ij}.
\label{eq:partial2}
\end{eqnarray}
In physical terms, the effect of the parameter $\alpha$ is to reduce the
contribution of long paths relative to short ones.  That is, for
$0<\alpha<1$, our similarity measure considers vertices to be more similar
if they have a greater than expected number of short paths between them,
than if they have a greater than expected number of long ones.  While this
is a natural route to take, it does mean we have introduced a new free
parameter into our calculations.  This seems a fair exchange: we have
traded the infinite number of free parameters in the expansion of
Eq.~\eref{eq:simsum} for just a single such parameter.  We discuss the
appropriate choice of value for $\alpha$ in Section~\ref{sect:alpha}.

The first term in Eq.~\eref{eq:partial2} is diagonal in the vertices $i$
and $j$ and hence affects only the similarity of vertices to themselves,
which we are not usually interested in, so we will henceforth drop it.
Thus, our final expression for the similarity is
\begin{equation}
S_{ij} = {2m\lambda_1\over k_ik_j} \biggl[ \biggl(
         \mat{I} - {\alpha\over\lambda_1} \mat{A} \biggr)^{-1}
         \biggr]_{ij}.
\label{eq:oursim1}
\end{equation}
Equivalently, we could write this in matrix form thus:
\begin{equation}
\mat{S} = 2m\lambda_1 \mat{D}^{-1} \biggl(
         \mat{I} - {\alpha\over\lambda_1} \mat{A} \biggr)^{-1}
         \mat{D}^{-1},
\label{eq:oursim2}
\end{equation}
where $\mat{D}$ is the diagonal matrix having the degrees of the vertices
in its diagonal elements: $D_{ij}=k_i\delta_{ij}$.

This similarity measure takes exactly the form we postulated in
Eq.~\eref{eq:matsim} with $\phi=\alpha/\lambda_1$, except for an overall
multiplier, which is trivial, and the leading factor of $1/k_ik_j$, which
is not.  This factor compensates for the fact that we expect there to be
more paths between pairs of vertices with high degree simply because there
are more ways of entering and leaving such vertices.  Its presence is
crucial if we wish to compare the similarities of vertex pairs having very
different degrees.

In practical terms, the calculation of the similarity matrix is most simply
achieved by direct multiplication.  Dropping the constant factor
$2m\lambda_1$ for convenience, we can rewrite Eq.~\eref{eq:oursim2} in the
form of Eq.~\eref{eq:initialSimExp} thus:
\begin{equation}
\mat{DSD} = {\alpha\over\lambda_1} \mat{A}(\mat{DSD}) + \mat{I}.
\label{eq:dsd}
\end{equation}
Making any guess we like for an initial value of $\mat{DSD}$, such as
$\mat{DSD}=0$, we iterate this equation repeatedly until it converges.  In
practice, for the networks studied here, we have found good convergence
after 100 iterations or less.

\subsection{Comparison with previous similarity measures}
\label{sect:previous}
Several other authors have proposed vertex similarity measures based on
matrix methods similar to ours~\cite{JW02,blondel:sim}.

Jeh and Widom~\cite{JW02} have proposed a method that they call
``SimRank,'' predicated, as ours is, on the idea that vertices are similar
if their neighbors are similar.  In our notation, their measure is
\begin{equation} 
S_{ij} = \frac{C}{k_ik_j} \sum_{u,v} A_{iu}A_{vj}S_{uv},
\label{eq:widom}
\end{equation}
where $C$ is a constant.  While this expression bears some similarity to
ours, Eq.~\eref{eq:initialSimExp}, it has an important difference also.
Starting from an initial guess for $S_{ij}$, one can iterate to converge on
a complete expression for the similarity, and this final expression
contains terms representing path counts between vertex pairs, as in our
case.  However, since the adjacency matrix appears twice on the right-hand
side of Eq.~\eref{eq:widom}, the expression includes \emph{only paths of
even length}.  This can make a substantial difference to the resulting
figures for similarity.  An extreme example would be a bipartite network,
such as a tree or a square lattice, in which vertices are separated either
only by paths of even length or only by paths of odd length.  In such
cases, those vertices that are separated only by paths of odd length will
have similarity \emph{zero}.  Even vertices that are directly connected to
one another by an edge will have similarity zero.  Most people would
consider this result counterintuitive, and our measure, which counts paths
of all lengths, seems clearly preferable.

Blondel~\etal~\cite{blondel:sim} considered similarity measures for
directed networks, i.e., based on asymmetric adjacency matrices, which is a
more complex situation than the one we consider.  However, for the special
case of a symmetric matrix, the measure of Blondel~\etal\ can be written as
\begin{equation}
S_{ij} = C \sum_{u,v} A_{iu}A_{vj}S_{uv},
\label{eq:blondel}
\end{equation}
where $C$ is again a constant.  This is very similar to the measure of Jeh
and Widom, differing only in the omission of the factor $1/k_ik_j$.  Like
the measure of Jeh and Widom, it can be written in terms of paths between
vertices, but counts only paths of even lengths, so that again vertices
separated only by odd paths (such as adjacent vertices on bipartite graphs)
have similarity zero.

\subsection{A measure of structural equivalence}
\label{sect:corollaries}
An interesting corollary of the theory developed in the previous sections
is an alternative measure of structural equivalence.  The structural
equivalence measures of Eq.~\eref{eq:structequiv} can be viewed as
similarity measures that count only the paths of length two between vertex
pairs; the number of common neighbors of two vertices is exactly equal to
the number of paths of length two.  Thus structural equivalence can be
thought of as just one term---the second-order term---in the infinite
series that defines our measure of regular equivalence.

The measures of Eq.~\eref{eq:structequiv} differ from one another in their
normalization.  The developments outlined in this paper suggest another
possible normalization, one in which we divide the number of paths of
length two by its expected value, Eq.~\eref{eq:pathsl2}, giving
\begin{equation}
\sigma = {2m\over k_ik_j} \left( {\av{k}\over\av{k^2}-\av{k}} \right)
         |\Gamma_i\cap\Gamma_j|.
\end{equation}
If we are concerned only with the comparative similarities of different
pairs of vertices within a given graph, then we can neglect multiplicative
constants and write
\begin{equation}
\sigma = \frac{|\Gamma_i\cap\Gamma_j|}{k_ik_j}
       = \frac{|\Gamma_i\cap\Gamma_j|}{|\Gamma_i|\,|\Gamma_j|}.
\end{equation}
This is, we feel, in many ways a more sensible measure of structural
equivalence than those of Eq.~\eref{eq:structequiv}.  It gives high
similarity to vertex pairs that have many common neighbors compared not to
the maximum number possible but to the \emph{expected} number of such
neighbors, and therefore highlights vertices that have a statistically
improbable coincidence of neighborhoods.  Of course, one could define
similar measures for paths of length 1 or 3 or any other length.  Or one
could combine all such lengths, which is precisely what our overall
similarity measure does.

\section{Tests of the method}
\label{sect:tests}
In this section we test our method on a number of different networks.  Our
first example is a set of computer-generated networks designed to have
known similarities between vertices.  In following sections we also test
the method against some real-world examples.

\subsection{Stratified model network}
\label{sect:stratified}
In many social networks, individuals make connections with others
preferentially according to some perceived similarity, such as age or
income.  Such networks are said to be \defn{stratified}, and stratified
networks present a perfect opportunity to apply our similarity measure:
ideally we would like to see that given only the network structure our
measure can correctly identify vertices that are similar in age (or
whatever the corresponding variable is) even when the vertices are not
directly connected to one another.

\begin{figure}[t]
\resizebox{\linewidth}{!}{\includegraphics{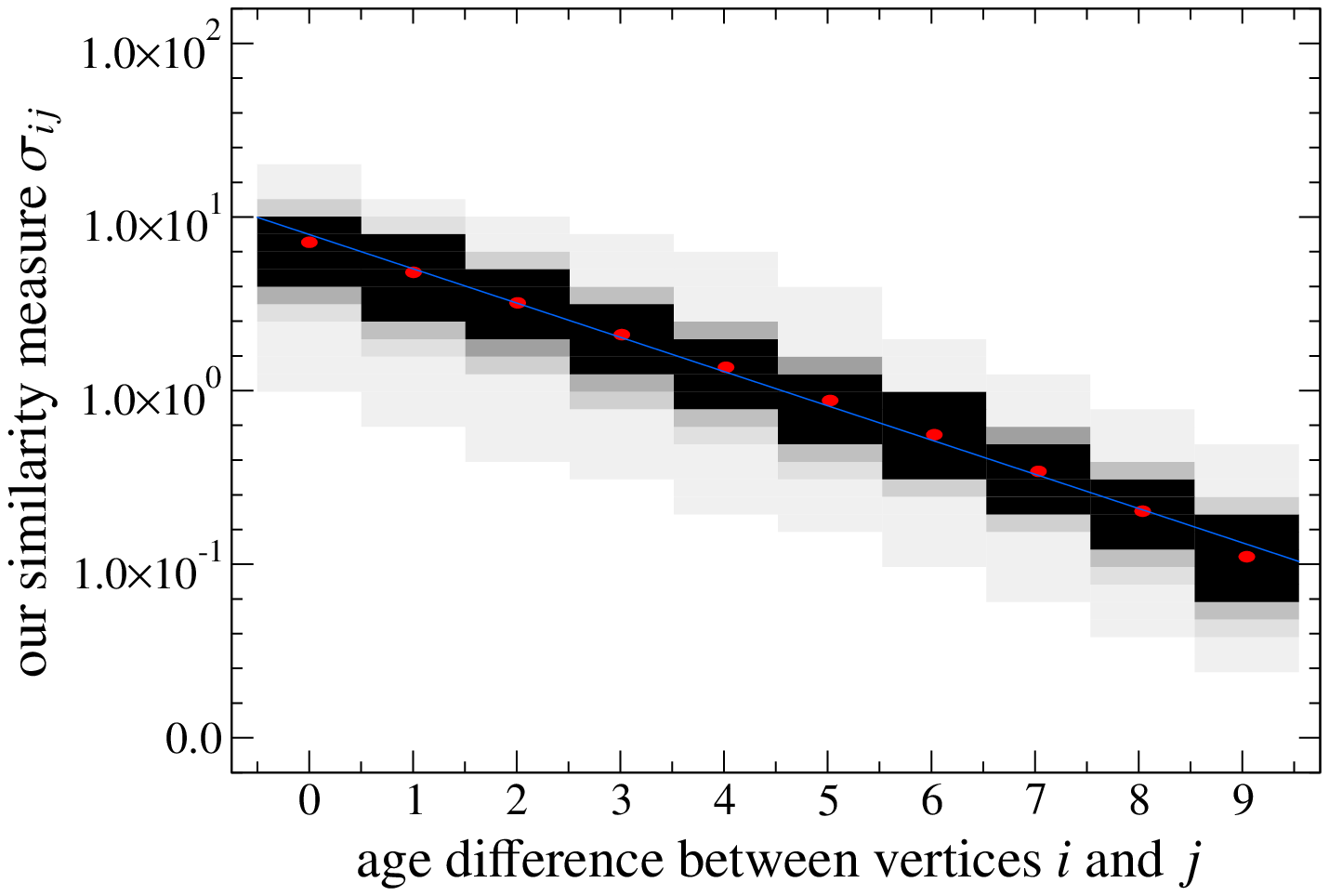}}\\
\resizebox{\linewidth}{!}{\includegraphics{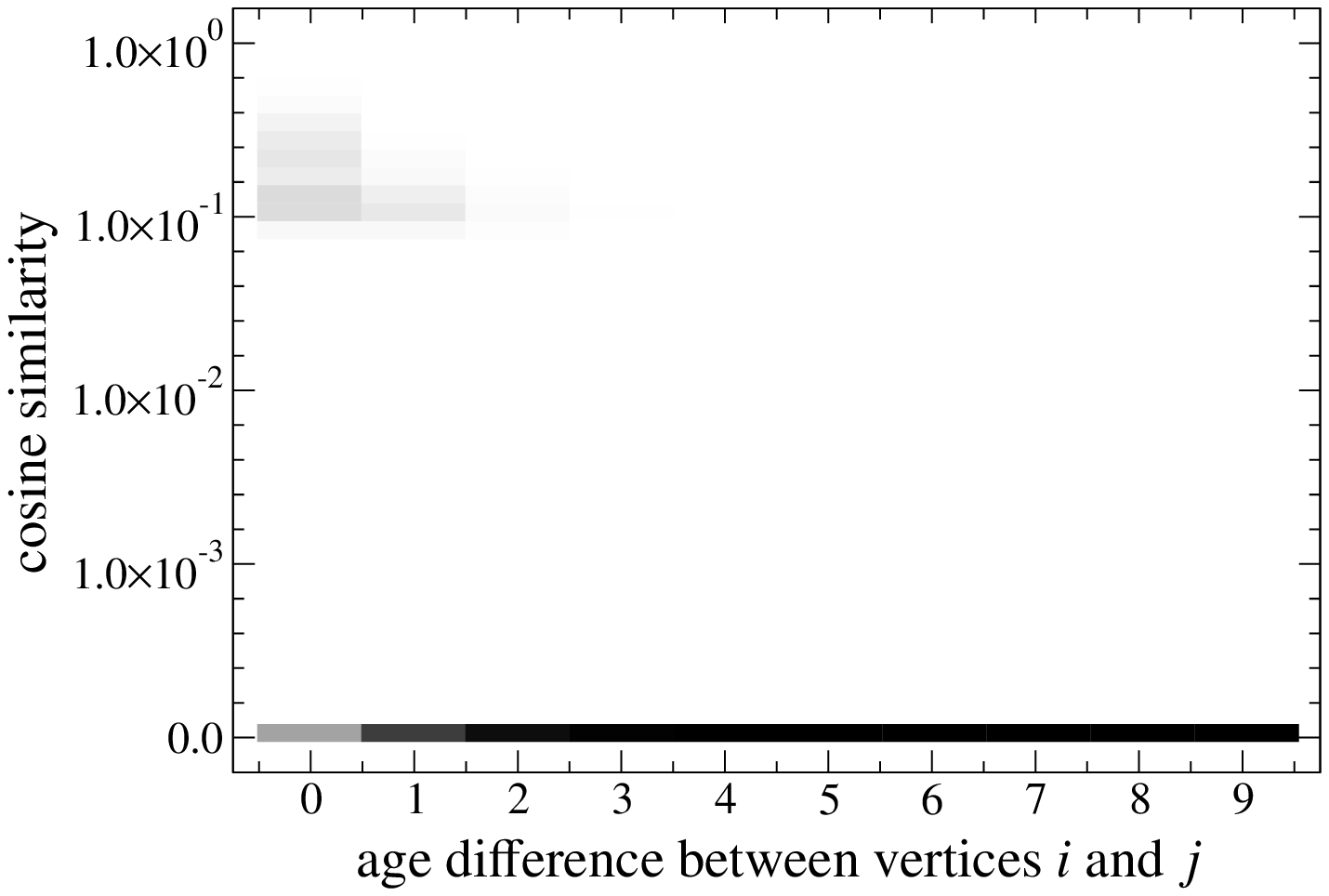}}
\caption{(a)~A density plot of the similarities of all vertex pairs not
directly connected by an edge in our stratified network model.  The points
give the average similarity as a function of age difference and the line is
a least-squares fit to a straight line.  (b)~A density plot of the cosine
similarity values for the same network.}
\label{fig:densityPlots}
\end{figure}

As a first test of our measure, we have created artificial stratified
networks on a computer.  Such networks offer a controlled structure for
which we believe we know the ``correct'' answers for vertex similarity.

In our model networks, each of $n=1000$ vertices was given one of ten
integer ``ages.''  Then edges were created between vertices with
probability
\begin{equation}
P(\Delta t) = p_0 e^{-a\Delta t},
\label{eq:pdeltat}
\end{equation}
where $\Delta t$ is the difference in ages of the vertices and $p_0$ and
$a$ are constants, whose values in our calculations were chosen to be
$p_0=0.12$ and $a=2.0$.  Thus the probability of ``acquaintance'' between two
individuals drops by a factor of $e^2$ for every additional year separating
their ages.

In order to calculate our similarity measure for this or any network we
need first to choose a value of the single parameter $\alpha$ appearing in
Eq.~\eref{eq:oursim1}.  In the present calculations we used a value
of~$\alpha=0.97$, which, as we will see, is fairly typical.  Since $\alpha$
must be strictly less than one if Eq.~\eref{eq:oursim1} is to converge,
$\alpha=0.97$ is quite close to the maximum.  We discuss in the following
section why values close to the maximum are usually desirable.

Figure~\ref{fig:densityPlots}a shows a density plot of the similarity
values for all vertex pairs in the model network not directly connected by
an edge, on semi-log scales as a function of the age difference between the
vertices.  The average similarity as a function of age difference is also
plotted along with a fit to the data.  We exclude directly connected pairs
in the figure because it is trivial that such pairs will have high
similarity and most of the interest in our method is in its ability to
detect similarity in nontrivial cases.

For comparison, we also show in Fig.~\ref{fig:densityPlots}b a density plot
of the cosine similarity, Eq.~\eref{eq:structequivb}, for the same network.
As the plot reveals, the cosine similarity is a much less powerful measure.
It is only possible for cosine similarity to be nonzero if there exists a
path of length two between the vertices in question.  Vertices with an age
difference of three or more rarely have such a path in this network and, as
Fig.~\ref{fig:densityPlots}b shows, such vertices therefore nearly all have
a cosine similarity of zero.  Thus cosine similarity finds only highly
similar vertices in this case and entirely fails to distinguish between
vertices with age differences between 3 and 9.  Our similarity measure by
contrast distinguishes these cases comfortably.

\subsection{Choice of $\alpha$}
\label{sect:alpha}
Our similarity measure, Eq.~\eref{eq:oursim1}, contains one free
parameter $\alpha$, which controls the relative weight placed on short and
long paths.  This parameter lies strictly in the range $0<\alpha<1$, with
low values placing most weight on short paths between vertices and high
values placing weight more equally both on short and long paths.  (Values
$\alpha>1$ would place more weight on long paths than on short, but for
such values the series defining our similarity does not converge.)

\begin{figure}[t]
\resizebox{\linewidth}{!}{\includegraphics{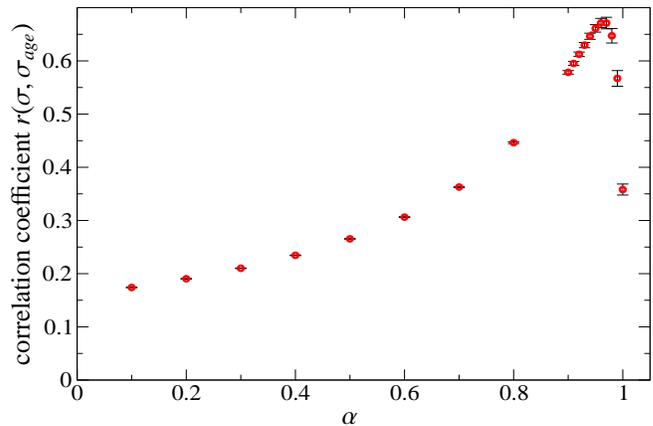}}
\caption{The correlation coefficient~$r(\sigma,\sigma_{age})$ for
correlation between our similarity measure and the probability of
connection, Eq.~\eref{eq:pdeltat}, in our stratified model, for a range of
values of $\alpha$.  The values given are averaged over an ensemble of
graphs generated from the model.  The maximum value is found to occur for
$\alpha\simeq0.97$.}
\label{fig:toyCorrCoeff}
\end{figure}

In order to extract quantitative results from our similarity measure we
need to choose a value for $\alpha$.  There is in general no unique value
that works perfectly for every case, but experience suggests some reliable
rules of thumb.  Our stratified network model, for instance, provides a
good guide.  Consider Fig.~\ref{fig:toyCorrCoeff}.  In this figure we
have  calculated the correlation coefficient of the similarity values
for vertex pairs determined using our method against the
probabilities, Eq.~\eref{eq:pdeltat}, of connections between the
vertices, which we consider to be a fundamental measure of vertices'
\textit{a priori} similarity.  As the figure shows, the
correlation is quite low for low values of~$\alpha$, but becomes strong as
$\alpha$ approaches one.  Only as $\alpha$ gets very close to one does the
correlation fall off again.  This appears to imply that a value of
$\alpha=0.9$ or greater should give the best results in this case.
Furthermore, it appears that, for values of $\alpha$ in this range, the
precise value does not matter greatly, all values around the maximum in the
correlation coefficient giving roughly comparable performance.

This we have found to be a good general rule: values of $\alpha$ close to
the maximum value of~1 perform the best, with values in the range 0.90 to
0.99 being typical.  Within this range the results are not highly sensitive
to the exact value.  We give another example to reinforce this conclusion
below.

The large typical values of $\alpha$ mean that paths of different lengths
are weighted almost equally in our similarity measure.  In other words, it
appears that our measure works best when long paths are accorded almost as
much consideration as short ones.  This contrasts strongly with structural
equivalence measures like the Jaccard index and the cosine similarity,
which are based exclusively on short paths---those of length two.  We
should be unsurprised therefore to find that our method gives substantially
better results than these older measures, as the example above shows.

\subsection{Thesaurus network}
\label{sect:thesaurus}
We now consider two applications of our method to real-world networks.  The
first is to a network of words extracted from a supplemented version of the
1911 US edition of \textit{Roget's Thesaurus}~\cite{roget}.  The thesaurus
consists of a five-level hierarchical categorization of English words.  For
example, the word ``paradise'' (level five) is cataloged under ``heaven''
(level four), ``superhuman beings and regions'' (level three), ``religious
affections'' (level two), and ``words relating to the sentient and moral
powers'' (level one).  Here we study the network composed of the 1000
level-four words, in which two such words are linked if one or more of the
level-five words cataloged below them are common to both.  For instance, 
the level-four words ``book'' and ``knowledge'' are connected because the
entries for both in the thesaurus contain the level-five terms ``book
learning'' and ``encyclopedia.''

In Table~\ref{tab:roget} we show the words most similar to the words
``alarm,'' ``hell,'' ``mean,'' and ``water,'' as ranked first by our
similarity measure and second by cosine similarity.  We used a value of
$\alpha=0.98$ in this case, on the grounds that this value gave the best
performance in other test cases (see below).

Since cosine similarity can be regarded as a measure of the number of paths
of length two between vertices, it tends in this example to give high
similarity scores for words at distance two in the thesaurus---synonyms of
synonyms, antonyms of synonyms, and so forth.  For example, cosine
similarity ranks ``pleasure'' as the word most similar to ``hell,''
probably because it is closely associated with hell's antonym ``heaven.''
By contrast, our measure ranks ``heaven'' itself first, which appears to be
a more sensible association.  Similarly, cosine similarity links ``water''
with ``dryness'', whereas our measure links ``water'' with ``plunge.''

\begin{table}
\setlength{\tabcolsep}{6pt}
\begin{tabular}{l|lr|lr}
  word & \multicolumn{2}{c|}{our measure} &
  \multicolumn{2}{c}{cosine similarity} \\
\hline
        & warning &  32.014 & omen & 0.51640\\
  alarm & danger & 25.769 & threat & 0.47141\\
        & omen & 18.806 & prediction & 0.34816\\
\hline
        & heaven & 63.382 & pleasure & 0.40825\\
  hell  & pain & 28.927 & discontent & 0.28868\\
        & discontent & 7.034 & weariness & 0.26726\\
\hline
        & compromise & 20.027 & gravity & 0.23570\\
  mean  & generality & 19.811 & inferiority & 0.22222 \\
        & middle & 17.084 & littleness & 0.20101 \\
\hline
        & plunge & 33.593 & dryness & 0.44721\\
  water & air & 25.267 & wind & 0.31623\\
        & moisture & 25.267& ocean & 0.31623
\end{tabular}
\caption{The words most similar to ``alarm,'' ``heaven,'' ``mean,'' and
``water,'' in the word network of the 1911 edition of \textit{Roget's
Thesaurus}, as quantified by our similarity measure and by the more
rudimentary cosine similarity of Eq.~\eref{eq:structequivb}.  For our
measure we used a value of $\alpha=0.98$ for the single parameter.}
\label{tab:roget}
\end{table}

\subsection{Friendship network of high school students}
\label{sect:friendship}
As a second real-world test of our similarity measure, we apply it to a set
of networks of friendships between school children.  The network data were
collected as part of the National Longitudinal Study of Adolescent Health
(AddHealth)~\cite{bear:addh}, and describe $90\,118$ students at 168
schools, including their school grade (i.e., year), race, and gender, as
well as their recent patterns of friendship.  It is well known that people
with similar social traits tend to associate with one
another~\cite{mcp:bird}, so we expect there to be a correlation between
similarity in terms of personal traits and similarity based on network
position.  This gives us another method for checking the efficacy of our
similarity measure.

The AddHealth data were gathered through questionnaires handed out to
students at 84 pairs of American schools, a school pair typically
consisting of one junior high school (grades 7 and 8, ages 12-14) and one
high school (grades 9-12, ages 14-18).  Here we look at a composite
of the school pairs with vertices from all six grades. Among other
things, the questionnaires circulated during the study asked
respondents to ``List your closest (male/female)
friends.  List your best (male/female) friend first, then your next best
friend, and so on.  (Girls/Boys) may include (boys/girls) who are friends
and (boy/girl) friends.''  For each of the friends listed the student was
asked to state which of five listed activities they had participated in
recently, such as ``you spent time with (him/her) last weekend''.  From
these answers a weight $w(i,j)$ is assigned to every ordered pair of
students $(i,j)$ such that $w(i,j)$ is 0 if $i$ has not listed $j$ as a
friend or 1 + the number of activities conducted otherwise.  From these
weights we construct an unweighted, undirected friendship network by adding
a link between vertices $i$ and $j$ if $w(i,j)$ and $w(j,i)$ are both
greater than or equal to a specified threshold value $W$.  As it turns out,
our conclusions are not very sensitive to the choice of $W$; the results
described here use $W=2$.

The networks so derived are not necessarily connected; they may, and often
do, consist of more than one component for each school studied.  To
simplify matters we here consider only on the largest component of each
network.  The largest component in some of the networks is quite small,
however, so to avoid finite size effects we have focused on networks of
more than 1000 students.

\begin{figure}[t]
\resizebox{\linewidth}{!}{\includegraphics{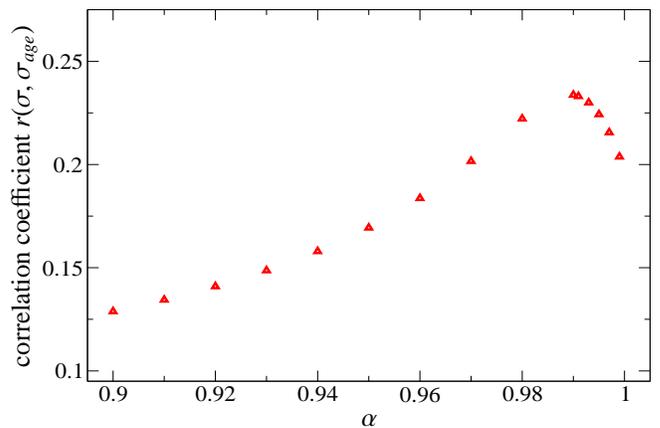}}
\caption{The correlation coefficient for correlation between our similarity
measure and the age difference of all vertex pairs in a single network, as
a function of~$\alpha$.  This plot is typical for the school networks
studied.}
\label{fig:addh}
\end{figure}

We first test our similarity measure using the method we used for the
stratified network of Section~\ref{sect:stratified}: we determine the
linear correlation coefficient between age difference (measured as
difference in grade) and our network similarity measure, for all vertex
pairs in a network.  We have calculated this correlation coefficient for a
range of values of~$\alpha$, the free parameter in our measure, and for a
selection of different networks.  The results for one particular network
are shown in Fig.~\ref{fig:addh}.  In this case the correlation coefficient
is maximized for $\alpha\simeq0.99$, which is again close to the maximum
possible value of~1.  For other networks we find maxima in the range from
0.96 to 0.99, which is in accord with the results of
Section~\ref{sect:alpha}.

These correlations between age difference and network similarity appear to
indicate that our similarity measure is able to detect some aspects of the
social structure of these networks.  To investigate this further, we have
taken the optimal values of $\alpha$ from the correlation coefficients and
used them to calculate the average similarity of vertex pairs that have a
known common characteristic, either grade or race, comparing that average
with the average similarity for vertex pairs that differ with respect to
the same characteristic.  The results are given in
Table~\ref{tab:addHealth}.

For school A the average similarity for pairs of students in the same grade
is a factor of eight greater than that for pairs in different grades---an
impressive difference.  It is possible, however, that this difference could
result purely from the contribution to the similarity from vertex pairs
that are directly connected by an edge.  It would come as no surprise that
such pairs tend to be in the same grade.  To guard against this, we give in
the fourth column of Table~\ref{tab:addHealth} results for calculations in
which all directly connected vertex pairs were removed.  Even with these
pairs removed we see that same-grade vertex pairs are on average
significantly more similar than pairs from different grades.

We have made similar calculations with respect to the race of students.
Students in school A did not appear to have any significant division along
racial lines (columns five and six of Table~\ref{tab:addHealth}), but this
school was almost entirely composed of students of a single race anyway, so
this result is not very surprising; it seems likely that the numbers were
just too small to show a significant effect.  School B was similar.
Schools C and D, however, show a marked contrast.  In school C, the average
similarity for students of the same race is a factor of five greater than
the average similarity for students of different races.  School C had a
population split 2:1 between two racial groups, in marked contrast with
schools A and B.  School D similarly appears to be divided by race,
although a little less strongly.  In this case there is a three-way split
within the population between different racial groups.  Possibly this more
even split with no majority group was a factor in the formation of
friendships between students from different groups.

\begin{table}
\setlength{\tabcolsep}{6pt}
\begin{tabular}{c|c|cccc}
  & & \multicolumn{4}{c}{similarity ratios}\\
  school & $n$ & \multicolumn{1}{c}{SG:DG} & \multicolumn{1}{c}{SG:DG*}
  & \multicolumn{1}{c}{SR:DR}
  & \multicolumn{1}{c}{SR:DR*} \\ \hline
  A & 1090 & 8.0 & 6.1 & 1.1 & 1.1 \\
  B & 1302 & 6.2 & 4.4 & 2.6 & 2.6 \\
  C & 1996 & 2.2 & 1.9 & 5.0 & 5.0 \\
  D & 1530 & 3.3 & 2.6 & 4.0 & 3.6
\end{tabular}
\caption{Network size $n$ and ratios of average similarity values
  for school networks in the AddHealth data set.  The column labeled SG:DG
  gives the ratio of average similarity for students in the same grade (SG)
  to average similarity for students in different grades (DG).  The column
  labeled SR:DR gives the ratio of average similarity for students of the
  same race (SR) to average similarity for students of different races
  (DR).  Columns marked with asterisks (*) give values of the same ratios
  but omitting vertex pairs connected directly by an edge.}
\label{tab:addHealth}
\end{table}

\section{Conclusions}
\label{sect:concs}
In this paper we have proposed a measure of structural similarity for pairs
of vertices in networks.  The method is fundamentally iterative, with the
similarity of a vertex pair being given in terms of the similarity of the
vertices' neighbors.  Alternatively, our measure can be viewed as a
weighted count of the number of paths of all lengths between the vertices
in question.  The weights appearing in this count are asymptotically equal
to the expected numbers of network paths between the vertices, which we
express in terms of the leading eigenvalue of the adjacency matrix of the
network and the degrees of the vertices of interest.  The resulting
expression for our similarity measure is given in Eq.~\eref{eq:oursim2}.

We have tested our measure against computer-generated and real-world
networks, with promising results.  In tests on computer-generated networks
the measure is particularly good at discerning similarity between vertices
connected by relatively long paths, an area in which more traditional
similarity measures such as cosine similarity perform poorly.  In tests on
real-world networks the method was able to extract sensible synonyms to
words from a network representing the structure of Roget's Thesaurus, and
showed strong correlations with similarity of age and race in a number of
networks of friendship among school children.  Taken together, these
results seem to indicate that the measure is capable of extracting useful
information about vertex similarity based on network topology.

The strength of similarity measures such as ours is their generality---in
any network where the function or role of a vertex is related in some way
to its structural surroundings, structural similarity measures can be used
to find vertices with similar functions.  For instance, similarity measures
can be used to divide vertices into functional
categories~\cite{rege:eco,rav:meta,wolfe:dots} or for functional prediction
in cases where the functionality of vertices is partly known ahead of
time~\cite{hh:pfp}.  We believe that the application of similarity measures
to problems such as these will prove a fruitful topic for future work.

\begin{acknowledgments}
This work was funded in part by the National Science Foundation under grant
number DMS--0405348, by the James S. McDonnell Foundation, and by a gift to
the University of Michigan from Robert D. and Janet E. Neary.
P.H. acknowledges support from the Wenner-Gren Foundation.

This work uses data from Add Health, a program project designed by
J. Richard Udry, Peter S. Bearman, and Kathleen Mullan Harris, and funded
by a grant P01-HD31921 from the National Institute of Child Health and
Human Development, with cooperative funding from 17 other agencies.
Special acknowledgment is due Ronald R. Rindfuss and Barbara Entwisle for
assistance in the original design.  Persons interested in obtaining data
files from Add Health should contact Add Health, Carolina Population
Center, 123 W. Franklin Street, Chapel Hill, NC 27516--2524
(addhealth@unc.edu).
\end{acknowledgments}


\end{document}